\documentclass[12pt]{article}

\usepackage[utf8]{inputenc}
\usepackage[T1]{fontenc}
\usepackage{amsmath, amssymb, amsfonts}
\usepackage{graphicx}
\usepackage{float}
\usepackage{enumitem}
\usepackage{parskip}
\usepackage{xcolor}
\usepackage{threeparttable}
\usepackage{siunitx}
\usepackage{hyperref}
\usepackage{geometry}
\usepackage{natbib}
\usepackage{booktabs}
\usepackage{caption}
\usepackage{subcaption}
\usepackage{chngcntr}
\usepackage{subcaption}

\definecolor{linkcolor}{RGB}{0, 90, 160}
\hypersetup{
    colorlinks=true,
    linkcolor=linkcolor,
    urlcolor=linkcolor,
    citecolor=linkcolor
}

\title{A Markov Chain Approach to Win Probability in Rugby Union}

\author{
  \begin{minipage}[t]{0.45\textwidth}
    \centering
    Kenny Watts \\
    University of Virginia \\
    \texttt{xen5ek@virginia.edu}
  \end{minipage}
  \hfill
  \begin{minipage}[t]{0.45\textwidth}
    \centering
    Jonathan Pipping-Gam\'on \\
    University of Pennsylvania \\
    \texttt{jpipping@wharton.upenn.edu}
  \end{minipage}
}
\date{October 2026}

\begin{document}

\maketitle

\begin{abstract}
Win probability tells us how likely a team is to win, but says little about how the rest of the match might play out. We develop a Markov chain model for rugby union that follows changes in score, time, field position, and possession, with transition probabilities estimated from phase-level match data. The same model lets us estimate win probabilities and expected points, calculate distributions for the next score and final margin, and simulate the remainder of a match from its current state. We assess predictive accuracy and calibration on 27 held-out matches after fitting the model to 105 matches from the same season. The model achieves similar test error to logistic regression and XGBoost, showing that we can model the progression of play while remaining competitive with methods that predict final outcomes directly.
\end{abstract}

\section{Introduction}

Markov models have been applied extensively in various sports including American football \citep{Goldner2012}, baseball \citep{Bukiet1997AMC}, and tennis \citep{newton2009monte}. Rugby union shares structural features with American football: variable field position, alternating possessions, and scoring plays with differing point totals. However, rugby union differs in that there aren't cleanly defined plays or drives, with continuous possessions that can vary greatly in length. This motivates a state-space model specifically built for rugby's unique structure.

We build a first-order Markov chain based on time, field position, and the current score.
The model supports both predictions of final outcomes and analysis of how games evolve between states.
We can calculate not just a win probability but a distribution over final margins and the next score, while also generating complete match trajectories through simulation.

In American football, win probability models are now well established and inform coaching decisions and the evaluation of play-level outcomes \citep{LockNettleton2014, YurkoVenturaHorowitz2019}. Early implementations relied on random forests and generalized additive models \citep{LockNettleton2014, YurkoVenturaHorowitz2019}, while more recent work has moved toward gradient-boosted methods capable of capturing nonlinear interactions between game-state variables \citep{Baldwin2021}. These models have also proven useful beyond in-game forecasting: game-state-based expected-value frameworks have been used to evaluate the rationality of coaching decisions, such as fourth-down strategy \citep{Romer2006}. 

Within rugby union itself, quantitative work has begun to adopt expected-points frameworks adapted from American football. \citet{Martinez-Arastey2025} lay methodological foundations for expected points in rugby union, with their phase-level data making it possible to construct the state space underlying our model. \citet{WattsPipping2026} applied an expected-points approach to a more specific in-game decision problem, evaluating whether kicking for goal or for touch is the better choice following a penalty. Both works demonstrate the value of state-based, points-oriented modeling in rugby, but neither develops a full generative state-space model capable of simulating complete match trajectories or producing dynamic, in-game win probability estimates alongside expected points.

\section{First-order Markov Chain}

A first-order Markov chain assumes that the distribution of the next state depends only on the current state. Let $S_n \in \mathcal{S}$ denote the state at time $n$, where $\mathcal{S}$ is the state space. Formally, for any timestep $n$, the next state $S_n$ is conditionally independent of all previous states $S_{n-2}, S_{n-3}, \ldots, S_1$ given the current state $S_{n-1}$:

\begin{equation}
P(S_n = s_{\text{new}} \mid S_{n-1} = s_{n-1}, \ldots, S_1 = s_1)
= P(S_n = s_{\text{new}} \mid S_{n-1} = s_{n-1})
\end{equation}

This property is often described as ``memoryless'': the system has no memory of how it arrived at its current state, only where it is now. In the context of our model, this means we assume that predicting what happens next requires only the current game situation (score differential, remaining time, pitch location, and possession) rather than the entire history of prior plays. Although certain path-dependent effects may exist, this assumption is both computationally efficient and reasonable given the difficulty of reliably inferring how specific play sequences influence future outcomes.

\section{Defining States}

A state $s \in \mathcal{S}$ is a tuple $s=(t,f,\delta,p)$ representing the game context at the start of a phase, where $t$ is the time bucket, $f$ is field position, $\delta$ is the exact score margin (the possession team's points minus its opponent's), and $p$ identifies the team in possession.
Transition probabilities are estimated for score buckets, which group margins into categories such as a lead of one to six points.
Together with the time and field-position buckets defined below, these give $6\times8\times7=336$ conditioning combinations.

\subsection*{Time bucket $\mathcal{T}$}
We divide each half into three segments based on seconds remaining, yielding $|\mathcal{T}| = 6$ buckets:
\[
  \mathcal{T} = \{\texttt{1h\_start},\ \texttt{1h\_middle},\ \texttt{1h\_end},\ \texttt{2h\_start},\ \texttt{2h\_middle},\ \texttt{2h\_end}\},
\]
where \texttt{start} and \texttt{end} correspond to the first and last two minutes of each half, respectively, and \texttt{middle} covers the remainder.

\subsection*{Field position $\mathcal{F}$}
Field position is expressed as the possession team's distance, in meters, from the opposition's try line. We discretize field position into $|\mathcal{F}| = 8$ ordinal zones centered on the following meter lines:
\[
  \mathcal{F} = \{2.5,\ 13.5,\ 31,\ 45,\ 55,\ 69,\ 86.5,\ 97.5\}.
\]

\subsection*{Score bucket $\mathcal{G}$}
The score margin is expressed from the perspective of the possession team---that is, the signed difference $\delta = \text{score}_{\text{poss}} - \text{score}_{\text{opp}}$---and coarsened into $|\mathcal{G}| = 7$ ordinal buckets:
\[
  g(\delta) =
  \begin{cases}
    \texttt{down\_14p}  & \text{if } \delta \leq -14, \\
    \texttt{down\_7\_13} & \text{if } -13 \leq \delta \leq -7, \\
    \texttt{down\_1\_6}  & \text{if } -6 \leq \delta \leq -1, \\
    \texttt{tied}        & \text{if } \delta = 0, \\
    \texttt{up\_1\_6}    & \text{if } 1 \leq \delta \leq 6, \\
    \texttt{up\_7\_13}   & \text{if } 7 \leq \delta \leq 13, \\
    \texttt{up\_14p}     & \text{if } \delta \geq 14.
  \end{cases}
\]

By conditioning on the possession team rather than a fixed home or away side, the transition dynamics are symmetric across teams by construction.

\section{Transition Matrix}

A first-order Markov chain is characterized by its transition probability matrix $\mathbf{P}$, where each entry $p_{s_n, s_{n+1}}$ represents the probability of transitioning from state $s_n$ to state $s_{n+1}$:
\[
  P(S_{n+1} = s_{n+1} \mid S_n = s_n) = p_{s_n,\, s_{n+1}}, \qquad s_n, s_{n+1} \in \mathcal{S}.
\]
In our rugby model, each row of $\mathbf{P}$ corresponds to a current game state $s_n\in\mathcal{S}$, and each column represents a possible next state or terminal match outcome.
Because scoring events can coincide with changes in possession and field position, transitions describe these changes jointly, together with the passage of time and the end of the half or match.

All transition probabilities are estimated empirically from the training matches, using the relative frequencies of joint transitions within each conditioning combination.
For combinations absent from training, we pool across score buckets at the same time and field position.

\section{Win Probability Estimation and Simulation}

\subsection{Direct Calculation of Outcome Probabilities}

To estimate win probabilities, we solve the absorption equations for the fitted Markov chain directly.
Let $\mathbf Q$ contain transition probabilities between ongoing match states, and let $\mathbf R$ contain probabilities of transitioning directly to a home win, draw, or away win.
The eventual outcome probabilities, $\mathbf B$, satisfy
\begin{equation}
\mathbf B=\mathbf Q\mathbf B+\mathbf R,
\qquad (\mathbf I-\mathbf Q)\mathbf B=\mathbf R.
\end{equation}
Solving this system accounts for the possible paths to each final outcome without Monte Carlo sampling error.\footnote{The calculation retains integer score margins from $-200$ to $200$ and tracks exits from this grid as a separate unresolved outcome. Its probability is below $3\times10^{-12}$ at all training and test evaluation states.}
The same fitted chain can also generate individual match trajectories, as illustrated below.

Consider a match state in the middle of the second half where the home team holds a 21-14 lead with possession at the 45-meter line. Given this combination of a 7-point advantage, possession, and favorable field position, we would expect the home team to have a strong likelihood of victory. The direct calculation gives the following outcome probabilities:

\begin{table}[h]
\centering
\caption{Directly Calculated Outcome Probabilities}
\begin{tabular}{ccc}
\hline
\textbf{Home Win} & \textbf{Away Win} & \textbf{Draw} \\
\hline
0.7836 & 0.1659 & 0.0505 \\
\hline
\end{tabular}
\label{tab:sim_results}
\end{table}

This approach allows us to estimate win probabilities from a given game state, accounting for the stochastic nature of rugby and the multiple pathways through which a match can unfold.

\subsection{Event Sequences}

A single simulation consists of a sequence of state transitions, where at each step the model samples the next transition according to the probabilities in $\mathbf{P}$. Consider the following excerpt from a simulation begun at kickoff:

\begin{table}[h]
\centering
\caption{Example Play-by-Play Sequence from a Single Simulation. Possession and scores are shown after each transition; points are signed relative to the team in possession before the transition.}
\resizebox{0.9\textwidth}{!}{%
\begin{tabular}{cllcccc}
\hline
\textbf{Play} & \textbf{Buckets Before} & \textbf{Points} & \textbf{Poss.} & \textbf{Home} & \textbf{Away} & \textbf{Buckets After} \\
\hline
9 & \texttt{1h\_middle\_31\_tied} & 0 & Away & 0 & 0 & \texttt{1h\_middle\_31\_tied} \\
10 & \texttt{1h\_middle\_31\_tied} & 0 & Away & 0 & 0 & \texttt{1h\_middle\_31\_tied} \\
11 & \texttt{1h\_middle\_31\_tied} & $+7$ & Away & 0 & 7 & \texttt{1h\_middle\_86.5\_up\_7\_13} \\
12 & \texttt{1h\_middle\_86.5\_up\_7\_13} & 0 & Home & 0 & 7 & \texttt{1h\_middle\_45\_down\_7\_13} \\
13 & \texttt{1h\_middle\_45\_down\_7\_13} & 0 & Away & 0 & 7 & \texttt{1h\_middle\_55\_up\_7\_13} \\
\hline
\end{tabular}
}
\label{tab:pbp_example}
\end{table}

Each transition can jointly change field position, possession, time bucket, and score. In this sequence, the away team scores 7 points, and possession changes on the following transition. Each transition is sampled according to the empirical probabilities observed in our training data for the corresponding conditioning combination.

\section{Win Probability: Performance and Error Reduction}

The dataset was obtained from a previously published study \citep{Martinez-Arastey2025}.

We evaluated the performance of our Markov chain model using the first 105 matches in recorded data order for training and the remaining 27 matches for testing. Transition probabilities were estimated using all 28,052 training phases. All models were evaluated on the same 2,479 first-phase observations (\texttt{Phase = 1}) in the test matches, using both team perspectives for a total of 4,958 observations.

We compared predictive accuracy against logistic regression, random forest, and XGBoost models, along with two baselines: a naive model that assigns 0.5 to each team regardless of game state, and a proportion-based model using the average home and away results in the training data. The comparison models use field position, score margin, card difference, possession, and score margin divided by the square root of seconds remaining plus one; the forest and XGBoost also include seconds remaining directly. Logistic regression uses a logit mean model, and the forest uses 250 trees. XGBoost stopping was selected using 84 training matches for fitting and 21 for validation, followed by refitting on all 105 training matches for the selected 88 rounds. The test matches were not used for fitting or model selection. Implementation details and fixed model settings are provided in the accompanying code.

\subsection{Performance Metrics}

For evaluation, a team's result is coded as $Y=1$ for a win, $Y=0.5$ for a draw, and $Y=0$ for a loss. The corresponding Markov prediction is the expected result,
\[
\widehat{\mu}=\widehat{P}(\text{Win})+\tfrac12\widehat{P}(\text{Draw}).
\]
Model accuracy is assessed using root mean squared error (RMSE) and mean squared error (MSE) between the predicted expected result and $Y$. Lower values indicate better predictive performance. We also calculate error reduction relative to the naive baseline:

\begin{equation}
\text{Error Reduction} = \frac{\text{MSE}_{\text{baseline}} - \text{MSE}_{\text{model}}}{\text{MSE}_{\text{baseline}}} \times 100\%
\end{equation}

Uncertainty in test-set comparisons is assessed using 2,000 bootstrap resamples of the 27 test matches, keeping all observations from a sampled match together and pairing the resamples across models. These intervals describe test-sample uncertainty with the fitted models held fixed.

\subsection{Results}

\begin{table}[H]
\centering
\caption{Model Performance and Error Reduction vs. Baselines (Out-of-Sample)}
\begin{tabular}{lcc}
\hline
\textbf{Model} & \textbf{MSE} & \textbf{Error Reduction} \\
\hline
XGBoost & 0.1834 & 23.1\% \\
Logistic Regression & 0.1817 & 23.7\% \\
\textbf{Markov Model} & \textbf{0.1832} & \textbf{23.1\%} \\
Random Forest & 0.2053 & 13.8\% \\
Home/Away Proportion Baseline & 0.2263 & 5.0\% \\
Naive Baseline & 0.2383 & -- \\
\hline
\end{tabular}
\label{tab:error_reduction}
\end{table}

The Markov chain model achieves a 23.1\% reduction in MSE relative to the naive baseline, performing competitively with XGBoost (23.1\%) and logistic regression (23.7\%) while retaining the ability to simulate complete game trajectories. The paired 95\% bootstrap intervals for the MSE difference between each alternative and Markov are $[-0.0091,\,0.0088]$ for XGBoost and $[-0.0087,\,0.0049]$ for logistic regression. These results do not establish a difference in predictive accuracy among the three models on the held-out matches.

\section{Model Calibration}

Beyond predictive accuracy, calibration assesses whether predictions align with observed outcomes. Here, a predicted expected result of 0.7 corresponds to an average result of 0.7, with draws counted as one half. We assess calibration using ten equal-width bins and compare the mean predicted expected result within each bin to the observed average result.

\begin{figure}[!htbp]
\centering
\begin{subfigure}[b]{0.46\textwidth}
    \centering
    \includegraphics[width=\textwidth]{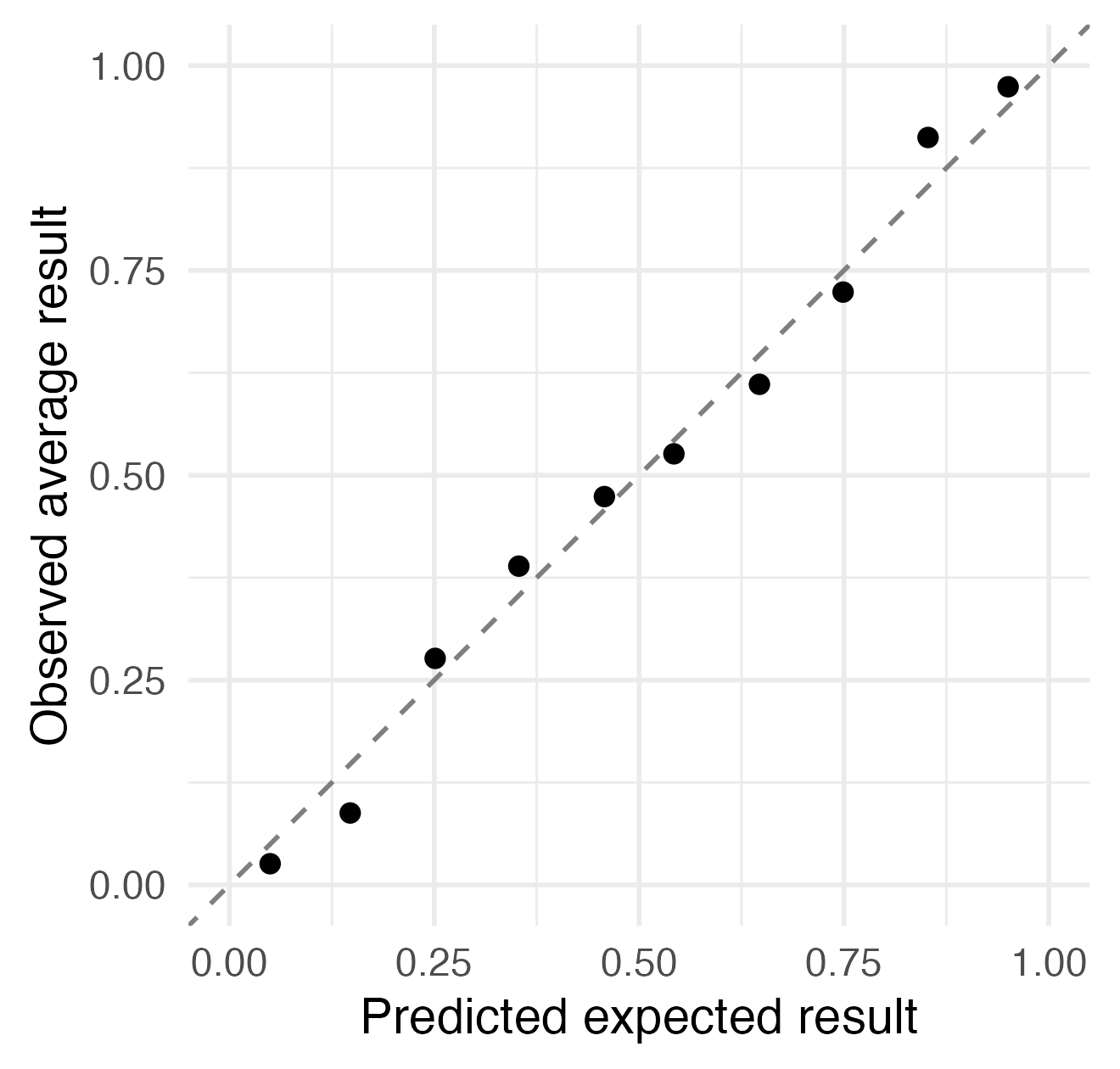}
    \caption{In-sample calibration}
    \label{fig:calib_in}
\end{subfigure}
\hfill
\begin{subfigure}[b]{0.46\textwidth}
    \centering
    \includegraphics[width=\textwidth]{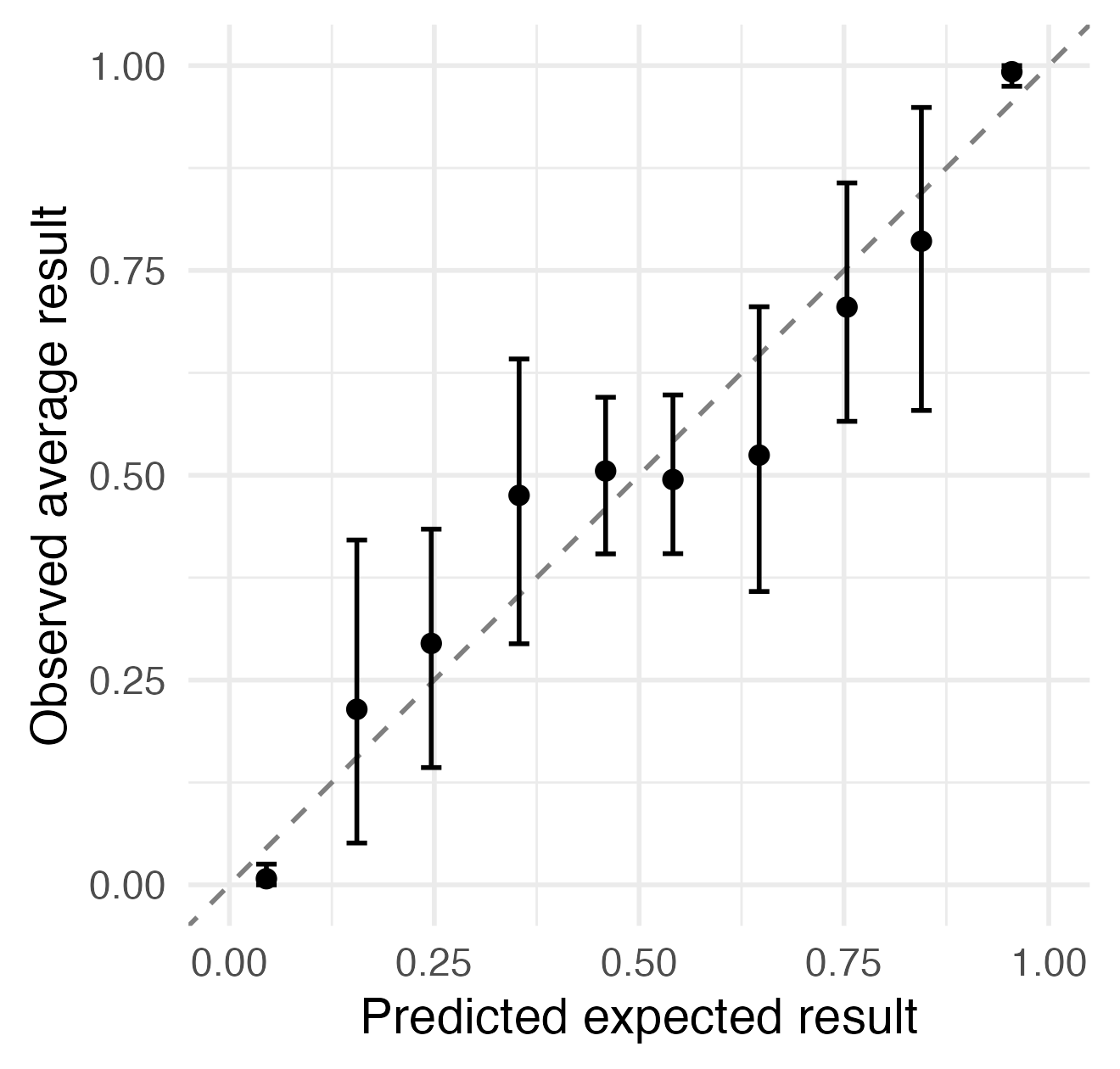}
    \caption{Out-of-sample calibration}
    \label{fig:calib_out}
\end{subfigure}
\caption{Calibration of the predicted expected result, $P(\mathrm{Win})+0.5P(\mathrm{Draw})$, against the observed result, coded as 1, 0.5, or 0. Both panels use first-phase observations and both team perspectives. Out-of-sample error bars show 95\% match-bootstrap intervals with the fitted model held fixed.}
\label{fig:calibration}
\end{figure}

Figure \ref{fig:calibration} shows calibration plots for the in-sample and out-of-sample datasets. The diagonal line represents agreement between predicted and observed average results. The out-of-sample intervals use the match bootstrap described above.

The out-of-sample results show deviations from the diagonal in some bins, with substantial uncertainty across the 27 test matches.
The match-bootstrap intervals reflect variation across test matches, with the fitted model held fixed.

\section{Beyond Win Probability}

\subsection{Next Score and Expected Points}

\begin{figure}[!htbp]
\centering
\begin{subfigure}[b]{0.46\textwidth}
    \centering
    \includegraphics[width=\textwidth]{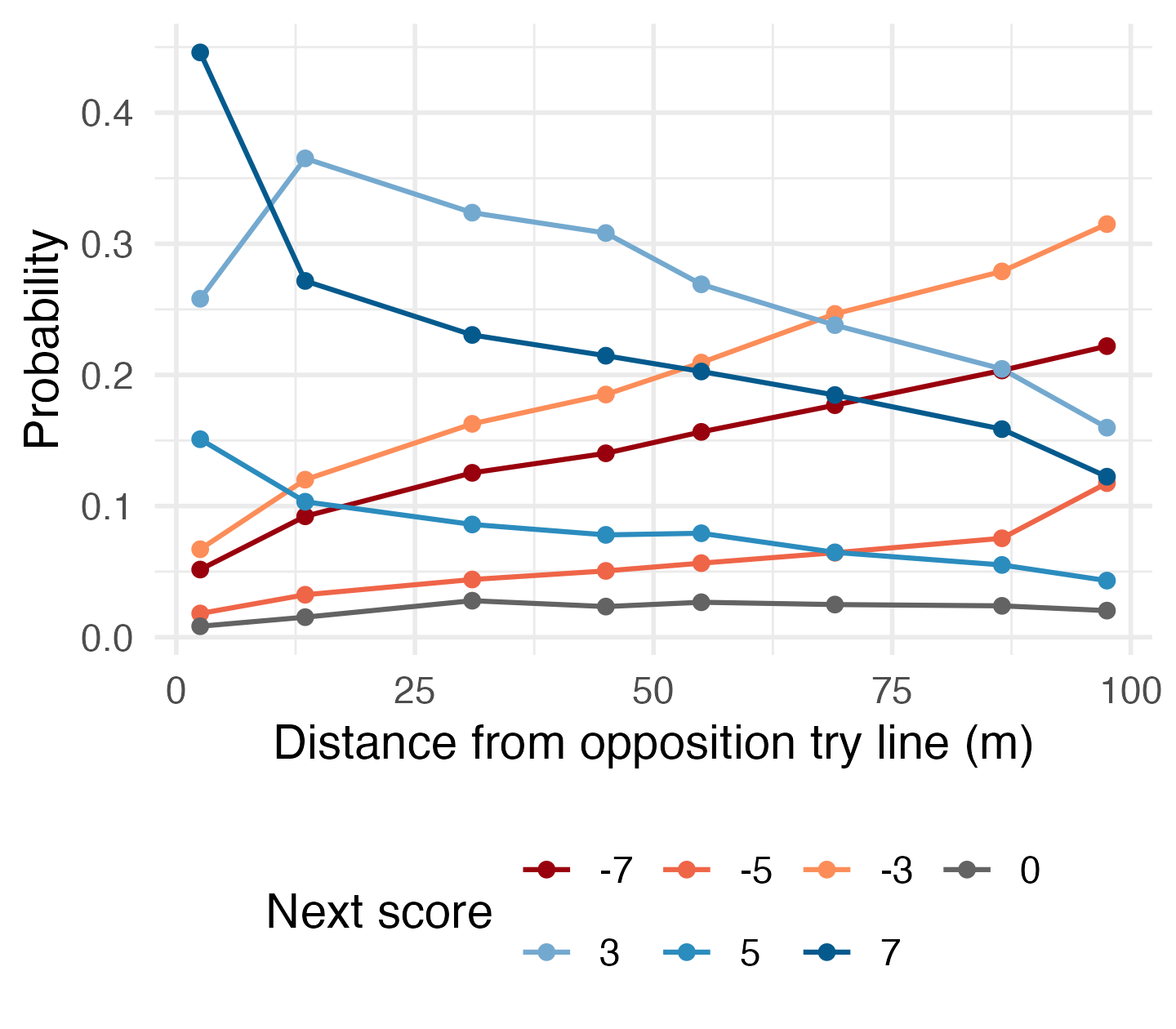}
    \caption{Next Score}
    \label{fig:next_score}
\end{subfigure}
\hfill
\begin{subfigure}[b]{0.46\textwidth}
    \centering
    \includegraphics[width=\textwidth]{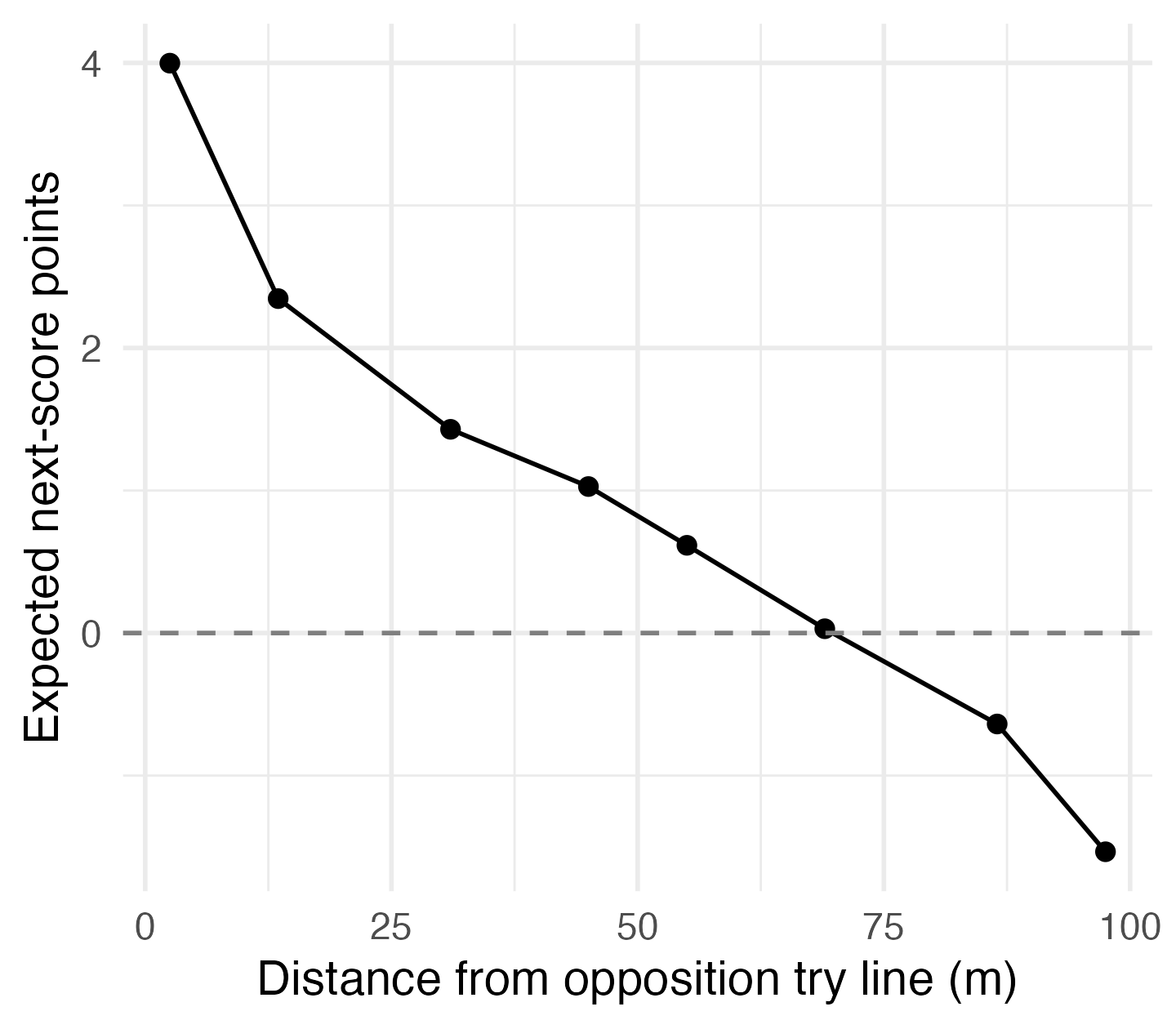}
    \caption{Expected Points}
    \label{fig:expected_points}
\end{subfigure}
\caption{Next-score probabilities and expected points with a tied score in the middle of the first half. Outcomes are measured relative to the team initially in possession, with zero representing no further score before halftime.}
\label{fig:score_analysis}
\end{figure}

The model can be used to value the current game state through the distribution of the next scoring event.
We consider the next score before the end of the current half, measured from the perspective of the team initially in possession.
The possible outcomes are $-7,-5,-3,0,3,5,7$, where zero indicates that the half ends without another score.
Figure~\ref{fig:next_score} shows these probabilities at different field positions, holding the score tied and the time bucket at the middle of the first half.
Figure~\ref{fig:expected_points} shows the weighted sum of these outcomes, giving the corresponding expected points.
Both are calculated directly from the fitted transition model.

\subsection{Final Score Margin}

\begin{figure}[!htbp]
\centering
\begin{subfigure}[b]{0.46\textwidth}
    \centering
    \includegraphics[width=\textwidth]{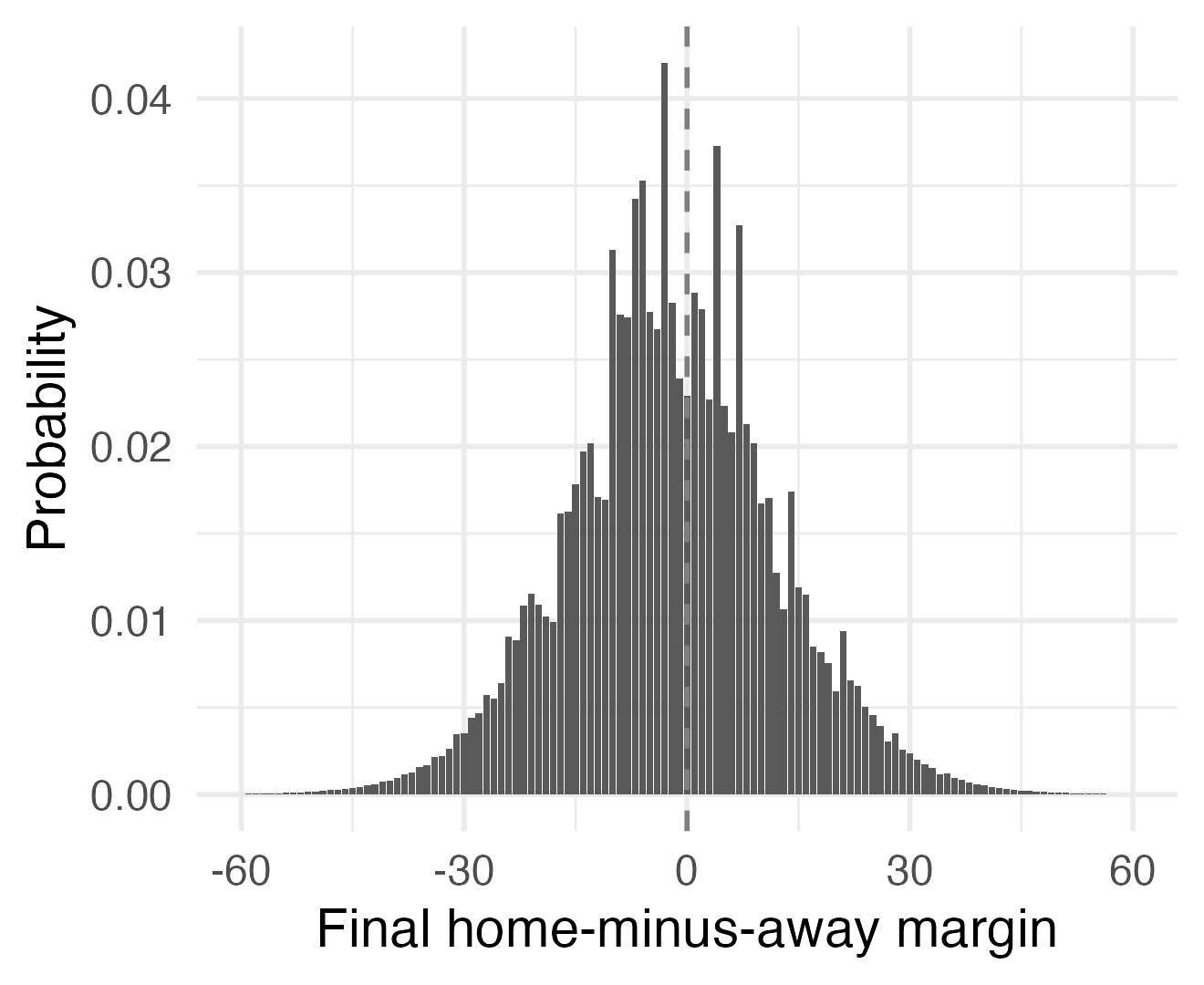}
    \caption{Final score margin given down 3 points in the middle of the first half}
    \label{fig:margin_first_half}
\end{subfigure}
\hfill
\begin{subfigure}[b]{0.46\textwidth}
    \centering
    \includegraphics[width=\textwidth]{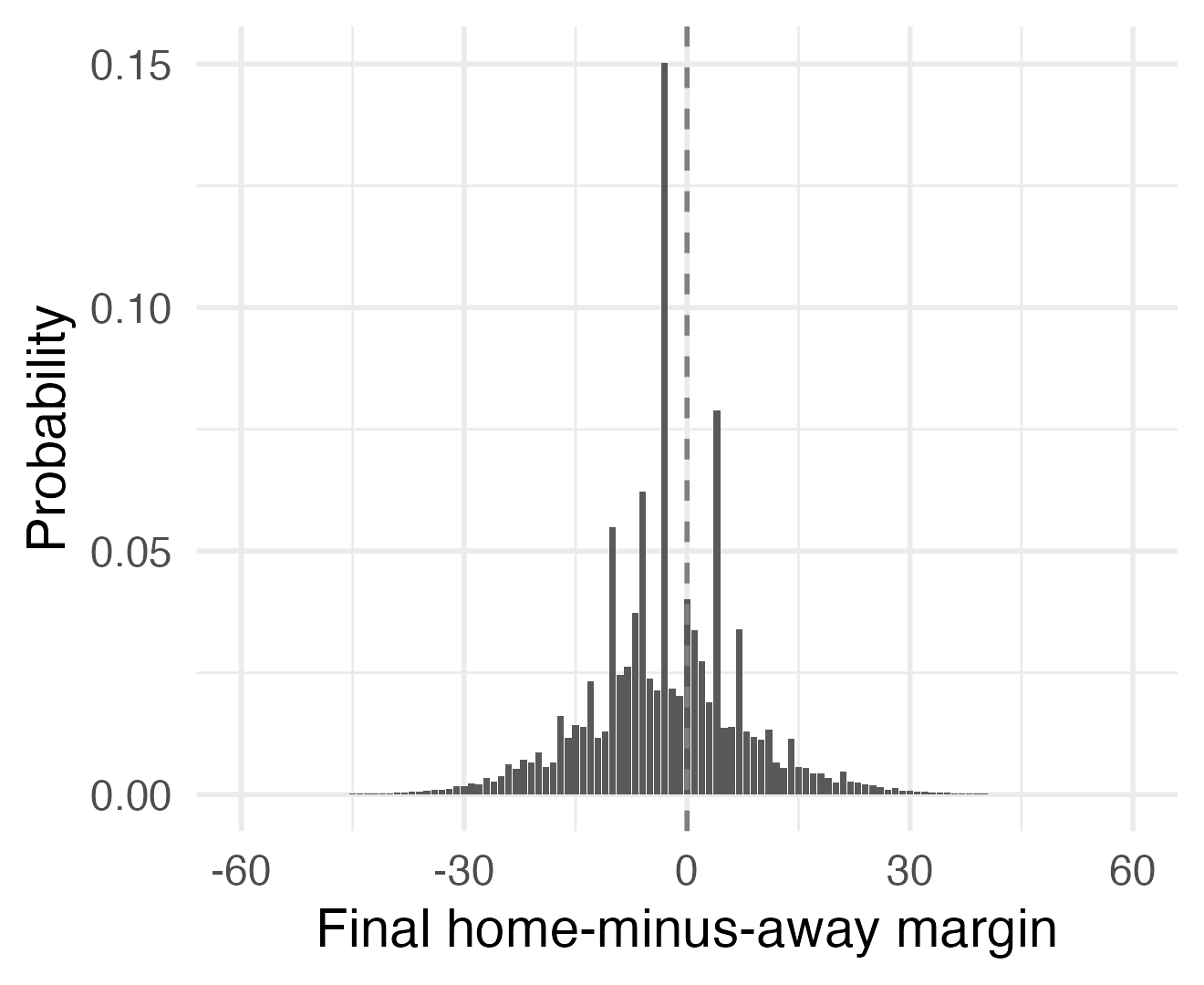}
    \caption{Final score margin given down 3 points in the middle of the second half}
    \label{fig:margin_second_half}
\end{subfigure}
\caption{Directly calculated final-margin distributions with the home team down 3 points and in possession at the 45-meter line. Both panels show margins from $-60$ to $60$; the omitted tail probability is below 0.0003 in each panel.}
\label{fig:margins}
\end{figure}

We can also analyze final scoring margins given a certain state, allowing us to answer questions such as ``if my team is down by X points, what is the probability we win the game by Y points?'' Figure~\ref{fig:margin_first_half} shows the final home score minus the final away score given that the home team is losing by 3 points in the middle of the first half, with possession at the 45-meter line. Figure \ref{fig:margin_second_half} shows the same state in the middle of the second half. Both distributions are calculated directly from the fitted chain, with terminal outcomes distinguished by final score margin.

The distribution of final margins is wider from the first-half state than from the second-half state, reflecting more opportunity for subsequent scoring under the fitted model. The current margin of $-3$ is the most likely final margin in both examples.

\section{Conclusion}

We developed a Markov chain model that constructs a transition matrix to estimate the likelihood of moving between game states. Beyond simply modeling these transitions, this approach allows teams to benchmark actual performance against expected outcomes, forecast final results, and trace how a game's trajectory unfolds over time.

More broadly, this framework offers a flexible, data-driven tool for in-game analysis that can be extended and refined with more data.
Our dataset consists of a single season with a total of 132 games.
This motivated the use of coarse time and scoring buckets.
Some transition estimates depend on few observations, while the middle-half buckets do not distinguish exact clock time.
Additional data could support finer states and more stable estimates.
Within this setting, the model performs competitively with the comparison models.

In terms of future work, team strength is not explicitly modeled; conditioning on the current score differential does not separately identify team strength. With multiple seasons of data, researchers could stratify matrices by team strength or adjust scoring transitions using offensive and defensive parameters for each team. Whether these additions improve calibration would require evaluation on held-out matches.

\bibliographystyle{apalike}
\bibliography{references}

\end{document}